\documentclass[conference]{IEEEtran}
\IEEEoverridecommandlockouts
\usepackage{cite}
\usepackage{amsmath,amssymb,amsfonts}
\usepackage{algorithmic}
\usepackage{graphicx}
\usepackage{textcomp}
\usepackage{xcolor}
\def\BibTeX{{\rm B\kern-.05em{\sc i\kern-.025em b}\kern-.08em
    T\kern-.1667em\lower.7ex\hbox{E}\kern-.125emX}}
\begin{document}

\title{Location Forensics Analysis Using ENF Sequences Extracted from Power and Audio Recordings
}

\author{\IEEEauthorblockN{1\textsuperscript{st} Dhiman Chowdhury, \textit{Student Member, IEEE}}
\IEEEauthorblockA{\textit{Electrical Engineering} \\
\textit{University of South Carolina}\\
Columbia, SC 29208, USA\\
dhiman@email.sc.edu}
\and
\IEEEauthorblockN{2\textsuperscript{nd} Mrinmoy Sarkar, \textit{Student Member, IEEE}}
\IEEEauthorblockA{\textit{Electrical and Computer Engineering} \\
\textit{North Carolina A \& T State University}\\
Greensboro, NC 27411, USA \\
msarkar@aggies.ncat.edu}
}

\maketitle

\begin{abstract}
Electrical network frequency (ENF) is the signature of a power distribution grid which represents the nominal frequency (50 or 60 Hz) of a power system network. Due to load variations in a power grid, ENF sequences experience fluctuations. These ENF variations are inherently located in a multimedia signal which is recorded close to the grid or directly from the mains power line. Therefore, a multimedia recording can be localized by analyzing the ENF sequences of that signal in absence of the concurrent power signal. In this paper, a novel approach to analyze location forensics using ENF sequences extracted from a number of power and audio recordings is proposed. The digital recordings are collected from different grid locations around the world. Potential feature components are determined from the ENF sequences. Then, a multi-class support vector machine (SVM) classification model is developed to validate the location authenticity of the recordings. The performance assessments affirm the efficacy of the presented work.
\end{abstract}


\section{Introduction}\label{sec1}
Location forensics analysis yields an important tool for anti-terrorist drives in regard to prevent and prosecute cyber crimes. Location-stamp verifications can be implemented investigating the variations of power system network frequency with respect to its fundamental value. Power system frequency is subject to instantaneous changes in accordance with load variations and control methodologies. Electrical network frequency (ENF) is the base frequency (50 or 60 Hz) of a power distribution system and ENF sequences are generated due to the fluctuations in frequency from the nominal ENF value. Since the ENF variations are uniform for a particular grid and are separable from grid-to-grid observation, these ENF sequences contain recognizable patterns of a power grid. When a multimedia signal (audio or video) is recorded close to a grid or directly from the power supply line, power signatures of that specific grid location are embedded into that recording due to the electromagnetic interference (EMI). Thereby, the multimedia recording can be applied for location forensics analysis in the situation, when the concurrent power recording is absent. Since ENF sequences carry the power signatures of a distribution grid, audio or video authenticity can be tested for location-stamp verification by extracting and analyzing the ENF signals of the recordings.

\begin{figure*}[t!]
	\includegraphics[height=2.5 in, width=7in]{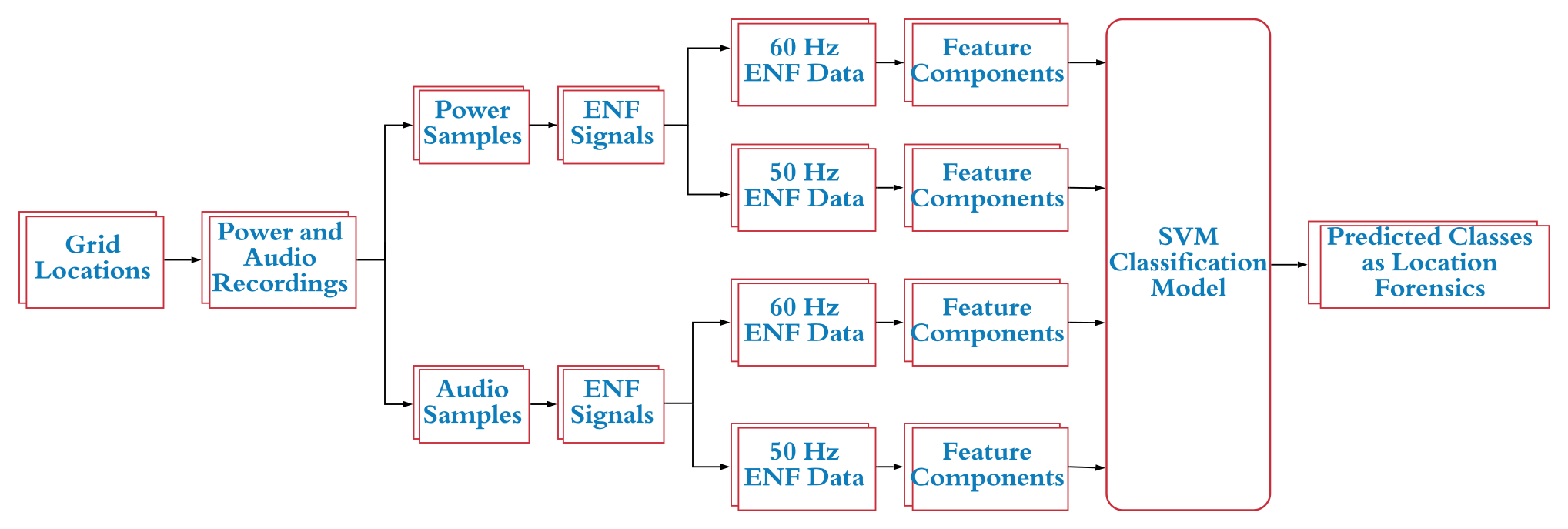}
	\caption{Work flow of the proposed location forensics analysis system} \label{fig111}
\end{figure*}

In this paper, ENF sequences are extracted from a number of power and audio signals recorded in different grid regions. Root MUSIC algorithm \cite{1} is used here to determine the ENF sequences. To maintain a moderate computational cost and complexity with considerably high precision, root MUSIC algorithm is applied in this work. After extracting ENF signals, potential feature vectors are obtained analyzing the statistical characteristics of the ENF patterns. Then, a multi-class support vector machine (SVM) classification model is trained and tested by these features to locate the regions of recordings. Thus, a complete location forensics validation framework based on inspecting power grid frequency variation trends is substantiated.

There are a number of works in which investigations of ENF signal recordings and characteristics and their applications in location-stamp verification are reported. A comprehensive ENF analysis of digital audio recordings for forensics and security applications is articulated in \cite{2}. Several novel methodologies on ENF extraction from power and multimedia signals are proposed in \cite{3} - \cite{10}. However, a number of location-stamp verification works based on ENF analyses are reported in \cite{11} - \cite{15}.

The proposed location forensic application system is developed and tested in MATLAB \textregistered\; and the training and testing accuracies to locate the regions of the power and audio recordings are obtained as 91.50 \% and 84.00 \% respectively. The major contributions of this research work are as follows.
\begin{itemize}
	\item Proposes potential and efficient feature components from extracted ENF sequences embedded in power and audio recordings captured from different grid locations.
	\item Implements a custom SVM classification model to localize the recordings effectively.
	\item Proposes a computationally cost-effective, simple and similar or more efficient location forensics analysis framework than those reported in \cite{11} - \cite{14}. 
\end{itemize}
This work significantly addresses the location-stamp validation research work documented in \cite{15}. This paper considers the same studied recordings database and ENF extraction methodology which are considered in \cite{15}. The extracted ENF signals are decomposed into low outliers and high outliers segments in \cite{15} before extracting feature vectors and taking into account those ENF segments, classification model is developed. The main difference between this paper and \cite{15} lies in defining the feature extraction domain to train the classifier. The proposed work may be less accurate than the work in \cite{15} to some extent, but it is computationally more time and cost effective and less complicated for practical applications.

The remainder of this manuscript is organized as follows. Section \ref{sec.2} describes the ENF extraction method and ENF database generation work flow. Section \ref{sec3} explains the proposed feature components. Section \ref{sec4} presents the developed classification model and its performance evaluations. Finally, Section \ref{sec.4} draws conclusion and future scope of the research paper.

\section{Location Specific ENF Determination and Database Generation} \label{sec.2}
\begin{table}[!b]
	\centering
	\caption{Location Forensics of the Training Dataset}
	\begin{tabular}{|p{1in}|p{1in}|p{1in}|}
		\hline
		Index & Grid Location & Notation \\
		\hline
		1 & Texas & $A$ \\
		\hline
		2 & Lebanon & $B$ \\
		\hline
		3 & Eastern U.S. & $C$ \\
		\hline
		4 & Turkey & $D$ \\
		\hline
		5 & Ireland & $E$ \\
		\hline
		6 & France & $F$ \\
		\hline
		7 & Tenerife & $G$ \\
		\hline
		8 & India & $H$ \\
		\hline
		9 & Western U.S. & $I$\\
		\hline \hline
		\multicolumn{3}{|p{3in}|}{\centering {*} 60 Hz Grids: $A$, $C$ and $I$}\\
		\multicolumn{3}{|p{3in}|}{\centering {**}50 Hz Grids: $B$, $D$, $E$, $F$, $G$ and $H$}\\
		\hline
	\end{tabular}%
	\label{table11}%
\end{table}%

There are power and audio recordings collected from nine different grids in the training dataset \cite{16}. The sampling frequency of the recordings is 1 kHz. There are three 60 Hz grids and six 50 Hz grids in the dataset. Table \ref{table11} presents the grid locations and the associated grid names of the training dataset. In the following, separation of the recorded power and audio signals, extraction of ENF sequences and generation of ENF database are documented. Fig. \ref{fig111} presents the systemic work flow of the proposed ENF based location authenticity verification framework.

\subsection{Power and Audio Signals Separation}
An inputted power or audio recording is segmented into a number of time frames. The window of each time frame is empirically taken as 5 min long. The dominant or center frequency $f_{d}$ of each signal component is determined using short time Fourier transform (STFT). Then, the signal to noise ratios (SNRs) are computed considering [$f_{d}-f_{b}$, $f_{d}+f_{b}$] as the power band of each signal. Here $f_{b}=0.5$ Hz is the step size of variation from the nominal frequency. Except this one, other bands are considered as noisy segments. The band power values are estimated using Welch power spectrum method. From the SNR values, the recordings are separated as power and audio signals. The conclusive derivation is that at the nominal frequency or harmonics, the SNR values of the power recordings are greater than those of the audio recordings.

\subsection{ENF Extraction}
ENF sequences are extracted from the disaggregated power and audio recordings applying the following computational process.

\subsubsection{Power ENF}
Each 5 min long power signal frame is processed through a 2nd order Butterworth band-pass filter designed with a frequency band of [40, 70] Hz. The filtered signal is then segmented into a number of time frames for ENF approximation. Each ENF time frame is empirically selected as 5 s long. Then, root MUSIC algorithm is applied to determine the ENF values.

\begin{figure}[t!]
	\includegraphics[width=3.5in]{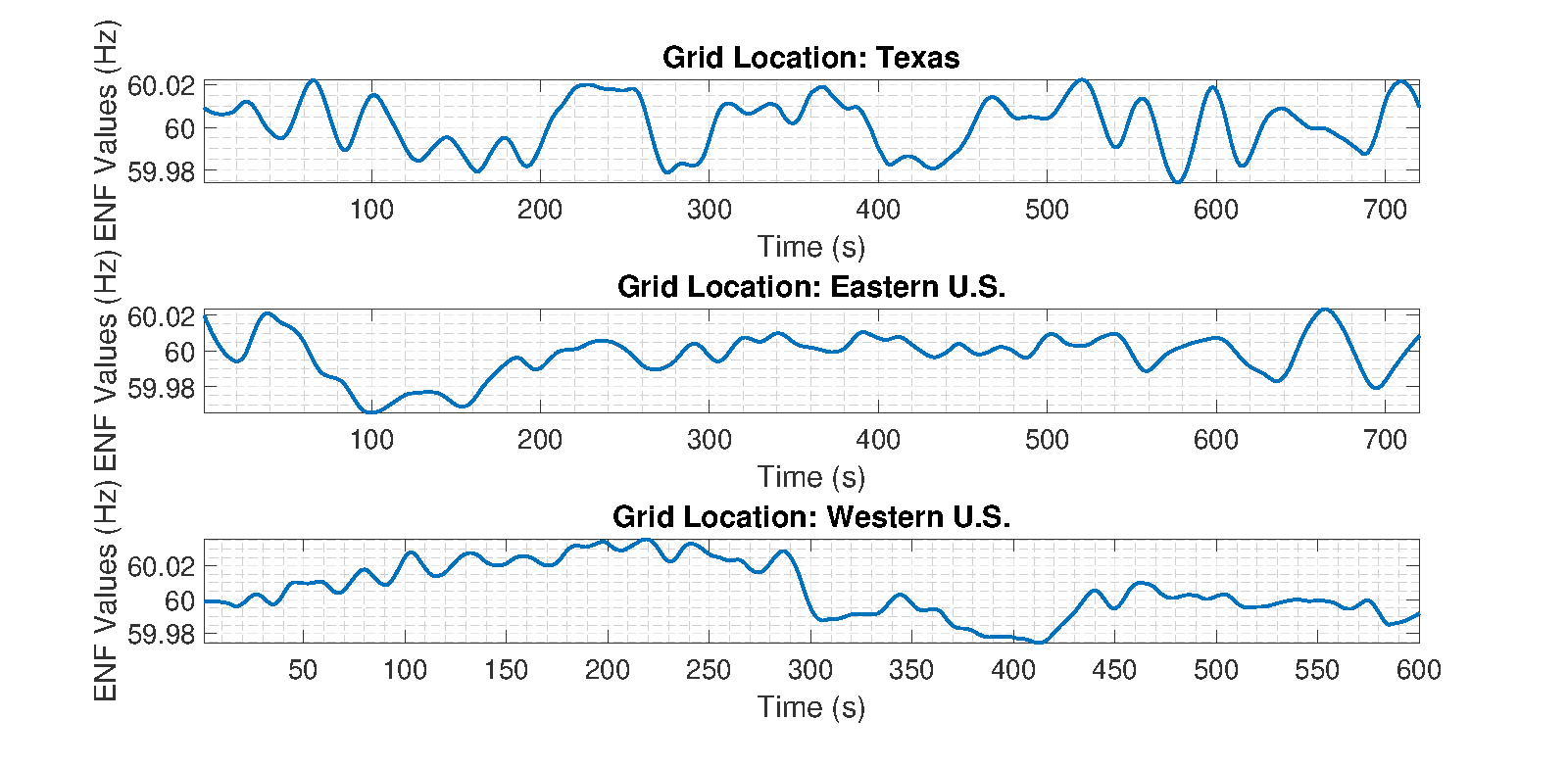}
	\caption{Sample ENF sequences embedded in power recordings captured from three 60 Hz grid locations} \label{fig1}
\end{figure}
\begin{figure}[t!]
	\includegraphics[width=3.5in]{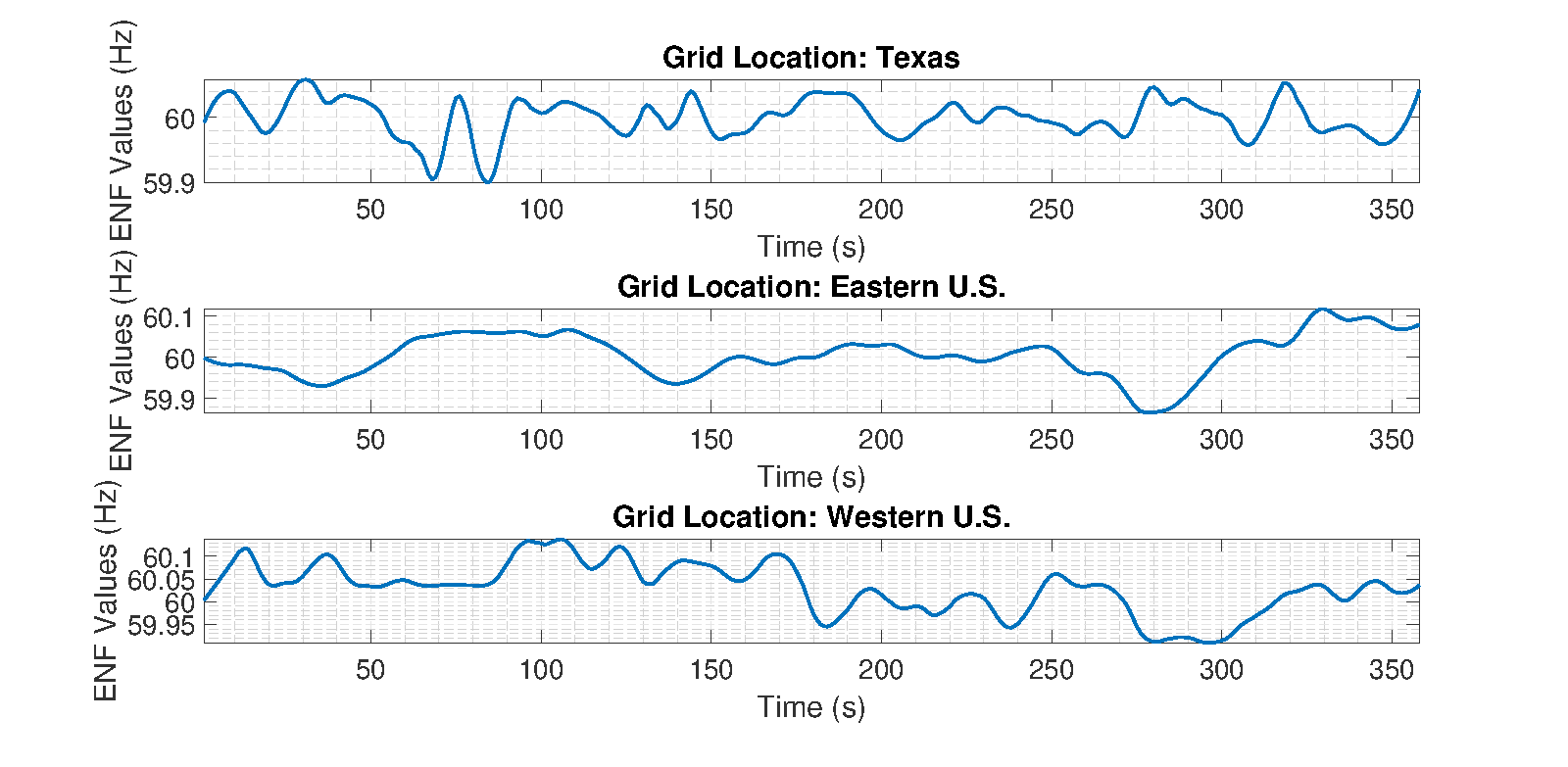}
	\caption{Sample ENF sequences embedded in audio recordings captured from three 60 Hz grid locations} \label{fig2}
\end{figure}
\begin{figure}[t!]
	\includegraphics[width=3.5in]{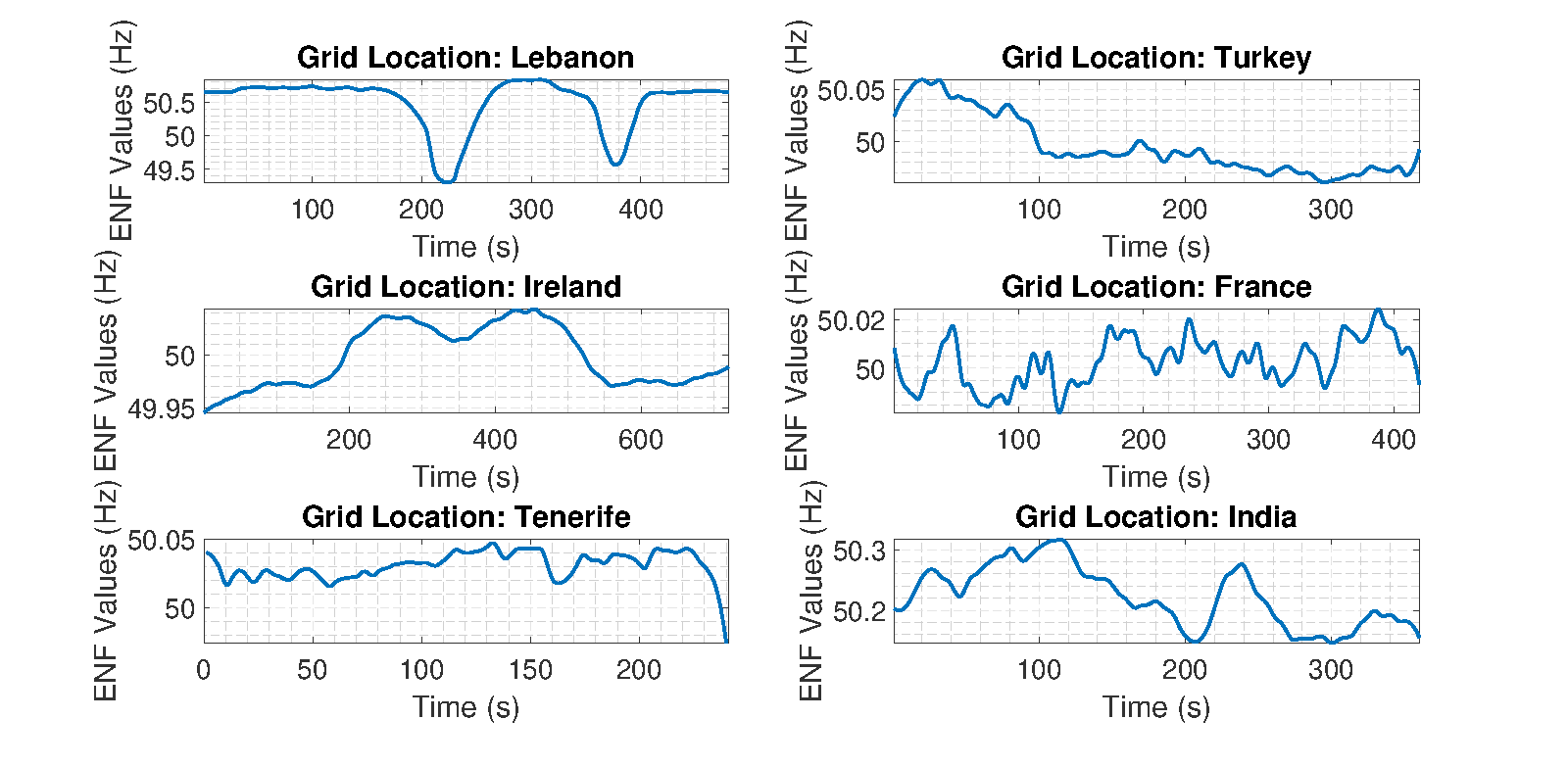}
	\caption{Sample ENF sequences embedded in power recordings captured from six 50 Hz grid locations} \label{fig3}
\end{figure}
\begin{figure}[t!]
	\includegraphics[width=3.5in]{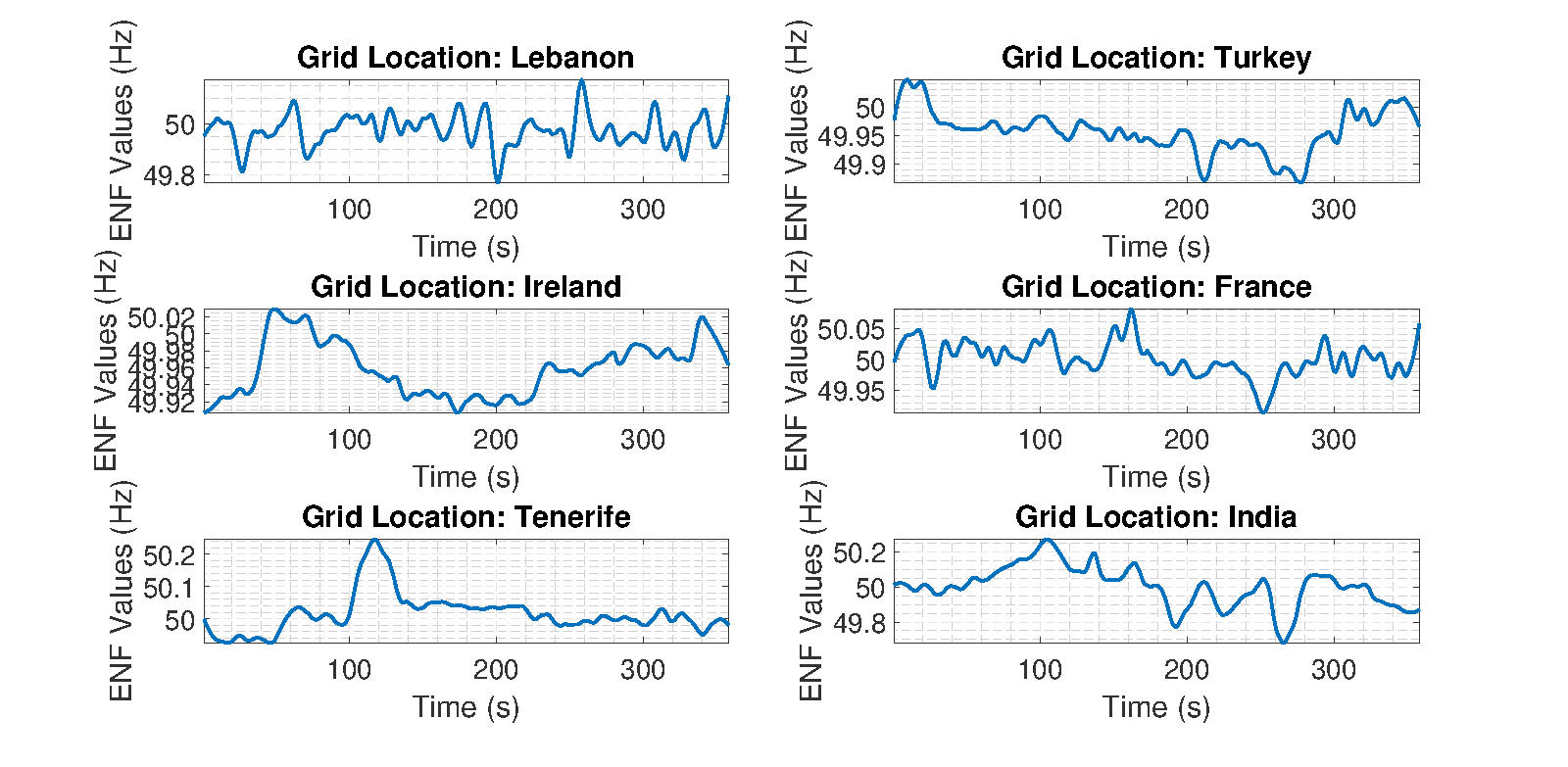}
	\caption{Sample ENF sequences embedded in audio recordings captured from six 50 Hz grid locations} \label{fig4}
\end{figure}

\subsubsection{Audio ENF}
Each 5 min long audio signal frame is processed through a 2nd order Butterworth band-pass filter designed with a frequency band of [$f_{d}-f_{b}$, $f_{d}+f_{b}$] Hz. The filtered signal is decomposed into 15 s long time frames, which overlap with each of the corresponding previous frames for 10 s. This implies that the overlapping of each frame with other is approximately 66.67 \%. In this follow-up, root MUSIC algorithm is applied to extract the ENF values.

Figs. \ref{fig1} - \ref{fig4} present the sample ENF sequences extracted from 50 Hz and 60 Hz power and audio recordings from different grid locations. The operating frequency fluctuations are subject to time to time load variations and overall power system control techniques. It is an accepted premise that loads change instantaneously with an accordance of energy demand and thereby the system frequency supposedly experiences random fluctuations. The fluctuations are random but are synchronized with load change such that if load increases, frequency goes below its base value and if load decreases, frequency goes up. If the control techniques are effective and
reliable, less fluctuations in frequency are observed which result in less instabilities in a distribution network. Therefore, it can be stated that ENF sequences measure the stability index of a power system.

From Fig. \ref{fig1}, it can be observed that the maximum ENF variations from the nominal 60 Hz value for power recordings captured from Texas, Eastern U.S. and Western U.S.
lie within the limits of [+0.02, -0.025] Hz, [+0.02, -0.03] Hz and [+0.03, -0.025] Hz respectively. For audio recordings collected from these three grids, the maximum ENF variations can be observed from Fig. \ref{fig2} lying within the limits of [+0.05, -0.1] Hz, [+0.11, -0.12] Hz and [+0.15, -0.1] Hz respectively. Similar observations can be made for 50 Hz power and audio ENF signals from Figs. \ref{fig3} and \ref{fig4}. From the approximated ENF fluctuation values, it can be implied that 60 Hz grids have better control methodology and more stable power system networks in comparison with 50 Hz grids, which are prone to large frequency variations during operation.

\begin{table*}[t!]
	\centering
	\caption{Extracted Feature Components from the ENF Sequences}
	\begin{tabular}{|p{1.55in}|p{1.55in}|p{1.55in}|p{1.55in}|}
		\hline
		\multicolumn{2}{|c|}{Power}  &
		\multicolumn{2}{c|}{Audio}
		\\ \hline
		\multicolumn{1}{|c|}{$f_{d}=60$ Hz} & \multicolumn{1}{c|}{$f_{d}=50$ Hz} & \multicolumn{1}{c|}{$f_{d}=60$ Hz} &
		\multicolumn{1}{c|}{$f_{d}=50$ Hz} \\ \hline
		1. Interquartile Range & 1. Mean & 1. Log Variance of Auto-Correlation Sequence & 1. Median \\
		\hline
		2. Log of Variance of Auto-Correlation Sequence & 2. Crest Factor & 2. Interquartile Range & 2. Power Spectral Density  \\
		\hline
		3. Log of Final Prediction Error $H$ of AR(4) Model & 3. Median & 3. Median & 3. 2nd Coefficient $G_{2}$ of AR(4) Model \\
		\hline
		- & 4. Waveform Length & 4. Modified Mean Absolute Value & 4. Log of Variance of Auto-Correlation Sequence \\
		\hline
		- & 5. Interquartile Range & - & - \\
		\hline
		- & 6. 2nd Coefficient $G_{2}$ of AR(4) Model & - & - \\
		\hline
	\end{tabular}%
	\label{table1}%
\end{table*}%

\subsection{ENF Database Generation}
From the extracted ENF sequences, 50 Hz and 60 Hz components are separated by measuring the mean or average values. Thus, for each power and audio recording 50 Hz and 60 Hz ENF signals are recognized separately and are stored in dedicated databases.

It is to be noted that the above described power and audio recordings separation, ENF sequences extraction and ENF database generation steps follow the approach reported in \cite{15}.

\section{Proposed Feature Components} \label{sec3}
This section describes the feature vectors extracted from the power and audio 50 Hz and 60 Hz ENF sequences. Statistics and basic signal processing techniques are employed here. Table \ref{table1} presents the extracted feature components from the ENF signals.

From experimental analysis, crest factor (CF) and interquartile range (IQR) are found to be potential feature functions. For an ENF signal, CF is measured as the ratio of the peak value to the root mean square (rms) value. IQR refers to the difference between the ENF value below which lie 25 \% of the entire sequence data, and that below which lie 75 \% of the entire sequence data. IQR analyzes the ENF sequence in terms of quartiles. Quartiles divide the data into four equal parts. The values that divide each part are called the first ($Q_{1}$), second ($Q_{2}$) and third ($Q_{3}$) quartiles. $Q_{1}$ is the middle value of the first half of the sequence. $Q_{2}$ is the median value and $Q_{3}$ is the middle value of the second half of the sequence. IQR is equal to $Q_{3} - Q_{1}$.

Welch power spectrum method is used to measure the power spectral density of an ENF sequence, which is proved to be a good feature component. However, 4th order autoregressive AR(4) model of an ENF sequence can be expressed as - 
\begin{equation}
f[k]=g_{1}f[k-1]+g_{2}f[k-2]+g_{3}f[k-3]+g_{4}f[k-4]+h
\end{equation}
Here $g_{1}$ - $g_{4}$ are the AR coefficients and $h$ is  the final prediction error (the variance estimate of the white noise input to the AR model). In this work, AR parameters are estimated using Burg method, where $g_{2}$ and log of $h$ are analyzed as potential feature components. Waveform length is a good candidate to extract a potential feature component $F_{c,WL}$, which is measured as -
\begin{equation}
F_{c,WL}=\sum\limits_{j=1}^{n-1}|f[j+1]-f[j]|
\end{equation}
Here $n$ is the sequence length. However,  Another potential feature vector $F_{c,MA}$ is derived from the modified mean absolute value function. It is defined as -
\begin{equation}
F_{c,MA}=\frac{\sum\limits_{j=1}^{n}0.5|f[j]|}{n}
\end{equation}
Extensive experiments are run to extract and select the promising feature components. Then Euclidean distance matrices are computed taking pairs of the features. The higher distance value of a particular feature from other features makes it
a better feature for selection. All other features those are extracted in the experiments such as 1st and 3rd coefficients of AR(4) model, kurtosis, skewness, mode, 5th and 6th order moments, r.m.s. shape factor, impulse factor and so forth have very small and inconsiderable Euclidean distances with
respect to the presented features. Therefore, those are not considered for final features to train the classification model.

\section{SVM Classification Model and Performance Evaluations} \label{sec4}
Based on the extracted feature components, a multi-class SVM classification model is developed. In this ``one-versus-one" classification approach, radial basis function (RBF) kernel is used. It is to be noted that the classifier algorithms are very much similar to those applied in \cite{15}.

The algorithm for training task is as follows.
\begin{itemize}
	\item After extracting feature components from the ENF signals, each feature vector $F_{c}$ is used in SVM classification algorithm as an input vector for training the prediction model.
	\item Table \ref{table2} presents the trained SVM models. There are 14 trained SVM models in total. 
\end{itemize}

\begin{table}[!t]
	\centering
	\caption{Trained SVM Models for 60 Hz and 50 Hz Recordings}
	\begin{tabular}{|p{0.65in}|p{0.65in}|p{0.65in}|p{0.65in}|}
		\hline
		\multicolumn{2}{|c|}{Power}  &
		\multicolumn{2}{c|}{Audio}
		\\ \hline
		$f_{d}=60$ Hz & $f_{d}=50$ Hz & $f_{d}=60$ Hz &
		$f_{d}=50$ Hz \\ \hline
		1. $AC$ & 1. $BF$ & 1. $AC$ & 1. $BF$ \\
		\hline
		2. $AI$ & 2. $HF$ & 2. $AI$ & 2. $DE$ \\
		\hline
		3. $CI$ & 3. $EF$ & 3. $CI$ & 3. $GH$ \\
		\hline
		- & 4. $DF$ & - & - \\
		\hline
		- & 5. $GF$ & - & - \\
		\hline
	\end{tabular}%
	\label{table2}%
\end{table}%

The algorithm for the testing purpose is as follows.
\begin{itemize}
	\item Let $S$ is the set of trained SVM models.
	\item Then, based on the signal type (audio or power) and $f_{d}$, the most appropriate model is pulled up from $S$. Let the pulled up model is $S_{j}$.
	\item Then, each feature vector $F_{c}$ is fed into $S_{j}$.
	\item The output of $S_{j}$ is the grid name (GN) associated with the input signal. Thereby, the input signal is classified as a particular GN. However, if the posterior probability of the predicted GN is less than a specified threshold value (0.6 in this case), then the output of the classifier is $NoG$. Class $NoG$ means `none of the grids' implying that the input is not a sample signal from any of the grids used for training.	
\end{itemize}

The ground truths of the testing dataset are available in \cite{16}. Table \ref{table3} presents the classification performance for different systems. In the training task, all 60 Hz power samples are classified correctly, whereas the 50 Hz power samples are recognized with more than 98 \% accuracy. In the testing scenario, 96 \% of 60 Hz power samples are classified correctly, whereas more than 95 \% of 50 Hz power samples are identified correctly. In the training casework, approximately 87 \% of 60 Hz audio signals are localized correctly, whereas around 83 \% of 50 Hz audio signals are classified correctly. In the testing scenario, more than 81 \% of 60 Hz audio samples are classified accurately, whereas more than 77 \% of 50 Hz audio samples are localized accurately. It can be observed that for both 60 Hz and 50 Hz data, power samples are localized more accurately than the corresponding audio samples. This outcome is somewhat expected here. For the overall training data, the system is 91.50 \% accurate, whereas for the overall testing data, the system is 84.00 \% accurate, which are considerably good results for real-time location forensic applications.

\begin{table}[t!]
	\centering
	\caption{Performance Evaluations in Terms of Training and Testing Accuracies (\%) of the Developed Classifier}
	\begin{tabular}{|p{0.27in}|p{0.27in}|p{0.27in}|p{0.27in}|p{0.27in}|p{0.27in}|p{0.27in}|p{0.27in}|}
		\hline
		\multicolumn{4}{|c|}{Power}  &
		\multicolumn{4}{c|}{Audio}
		\\ \hline
		\multicolumn{2}{|c|}{$f_{d}=60$ Hz} & \multicolumn{2}{c|}{$f_{d}=50$ Hz} & \multicolumn{2}{c|}{$f_{d}=60$ Hz} &
		\multicolumn{2}{c|}{$f_{d}=50$ Hz} \\ \hline
		Train. & Test. & Train. & Test. & Train. & Test. & Train. & Test. \\
		\hline
		100.00 & 96.00 & 98.33 & 95.50 & 87.05 & 81.66 & 83.25 & 77.27 \\
		\hline
		\cline{2-8}\hline \hline
		\multicolumn{4}{|c|}{Power \& Audio Training} & \multicolumn{4}{c|}{Power \& Audio Testing} \\
		\hline
		\multicolumn{4}{|c|}{\textbf{91.50}} & \multicolumn{4}{c|}{\textbf{84.00}} \\
		\hline	
	\end{tabular}%
	\label{table3}%
\end{table}%

\section{Conclusion \& Future Work} \label{sec.4}
Location forensics analysis is an important tool for security applications in the modern world. Different types of criminal and anti-social activities can be prevented and prosecuted by using location-stamp information of digital recordings. Therefore, novel and reliable location authenticity verification methods to investigate power and multimedia signals are significant.

In this paper, an efficient location forensics analysis method based on ENF signals of power and audio recordings captured from different grid locations around the world is presented. Potential feature components are extracted from the ENF sequences and a multi-class SVM classifier is developed to locate the regions of recordings. The obtained locations are verified with ground truths of the testing samples and the performance evaluations underscore the reliability of the proposed work.

The future scope of the presented ENF based location authenticity system is to employ the research methodologies for video recordings in absence of concurrent power grid signals. In addition, more robust ENF extraction technique can be conceptualized and used for real-time security applications.

\end{document}